\documentclass{article}
\usepackage{spconf,amsmath,graphicx}

\usepackage{amssymb}

\usepackage{tikz}
\usepackage{pgfplots}
\pgfplotsset{compat=1.3}
\usetikzlibrary{arrows,snakes,shapes}

\usepackage{color}
\DeclareMathAlphabet{\mathbit}{OML}{cmr}{bx}{it}

\DeclareMathOperator{\Q}{Q}
\DeclareMathOperator{\E}{E}

\newcommand\Sign{\operatorname{sign}}

\DeclareMathOperator{\T}{\operatorname{T}}

\DeclareMathOperator{\fieldR}{\mathbb{R}}

\newcommand{\ve}[1]{\boldsymbol{#1}}

\newcommand{\exdi}[2]{\E_{#1} \left[#2\right]}

\newcommand{\expb}[1]{\exp{\left(#1\right)}}

\newcommand{\asympequal}{\stackrel{\text{a}}{=}}

\newcommand{\sign}[1]{\Sign{\left(#1\right)}}

\newcommand{\qfunc}[1]{\Q \left(#1\right)}
\newcommand{\qfuncptwo}[1]{\Q^2 \left(#1\right)}
\DeclareMathAlphabet\mathbfcal{OMS}{cmsy}{b}{n}

\title{Performance Analysis for Pilot-based 1-bit Channel Estimation\\with Unknown Quantization Threshold}
\name{
Manuel Stein$^\dagger$, 
Shahar Bar$^\star$, 
Josef A. Nossek$^\dagger$, 
and Joseph Tabrikian$^\star$
\thanks{This research was partially supported by the Israeli Science Foundation (ISF, Grant 1392/11) and by the Yaakov ben Yitzhak Hacohen scholarship.}
}
\address{
 $^\dagger$Institute for Circuit Theory and Signal Processing, Technische Universit\"at M\"unchen, Germany\\
$^\star$Department of Electrical and Computer Engineering,  Ben-Gurion University of the Negev, Israel\\
E-mail: manuel.stein@tum.de, shahba@post.bgu.ac.il, josef.a.nossek@tum.de, joseph@ee.bgu.ac.il
}

\begin{document}
\maketitle
\begin{abstract}
Parameter estimation using quantized observations is of importance in many practical applications. Under a symmetric $1$-bit setup, consisting of a zero-threshold hard-limiter, it is well known that the large sample performance loss for low signal-to-noise ratios (SNRs) is moderate ($\frac{2}{\pi}$ or $-1.96$dB). This makes low-complexity analog-to-digital converters (ADCs) with $1$-bit resolution a promising solution for future wireless communications and signal processing devices. However, hardware imperfections and external effects introduce the quantizer with an unknown hard-limiting level different from zero. In this paper, the performance loss associated with pilot-based channel estimation, subject to an asymmetric hard limiter with unknown offset, is studied under two setups. The analysis is carried out via the Cram\'{e}r-Rao lower bound (CRLB) and an expected CRLB for a setup with random parameter. Our findings show that the unknown threshold leads to an additional information loss, which vanishes for low SNR values or when the offset is close to zero.
\end{abstract}
\section{Introduction}
\label{sec:intro}
\indent Due to the fact that the complexity of a sampling device scales exponentially $\mathcal{O}(2^b)$ with the number of bits $b$ which are used for the representation of the digital amplitude values, the analog-to-digital converter (ADC) has been identified as a crucial part for the design of hardware and energy-efficient signal processing systems \cite{Walden99}. Thus, from a hardware-aware perspective, ADCs with $1$-bit output are an attractive design option. On the other hand, the highly nonlinear transformation associated with hard-limiting causes a substantial information loss during the transition from the analog to the digital domain. Fortunately, the performance loss for low SNR is $\frac{2}{\pi}$ or equivalently $-1.96$ dB, and therefore moderate \cite{Vleck66}. Also, using a higher sampling rate \cite{Koch10,Krone12} or adjustment of the analog radio front-end \cite{SteinWCL15} allows compensation for the quantization loss. In addition, the optimization of the $1$-bit ADC itself has been studied. The work of \cite{Balkan10} shows how to set the hard-limiting threshold in order to maximize the Fisher information, while \cite{Koch13} focuses on communication rates and studies the Shannon information with $1$-bit quantization and asymmetric known offsets at the receiver. Another line of works, e.g. \cite{Dabeer06_2,Zeitler12}, investigates randomization of the quantization level, i.e., applying dithering before the quantizer.\\
\indent A problem that arises when the quantization level of a $1$-bit ADC is to be adaptively controlled, is the fact that a high resolution digital-to-analog converter (DAC) is required in order to set an analog offset voltage source. As also the complexity of DACs scales $\mathcal{O}(2^b)$ with the number of bits $b$, this stands in contradiction with the main motivation behind $1$-bit ADC technology, which is an energy and hardware efficient radio front-end. Therefore, low-complexity $1$-bit ADCs will lack the feature of an accurately adjustable quantization level. Rather, a low-cost sampling device will be constructed such that the hard-limiting level is fixed to a constant value. Inevitable mismatches of the circuit parts during the production process and external effects will lead to an unknown quantization level of the sampler during runtime.
\\
\indent In this paper, the performance loss associated with $1$-bit quantization and an unknown threshold is analyzed for the application of pilot-based channel estimation. The problem is first studied under the assumption that the channel parameter and quantization level are deterministic unknown. Then, a hybrid setup is considered, where the channel parameter is subject to a (known) prior distribution, while the hard-limiting offset is deterministic. The considered ADC allows to operate at high rates and provides binary data that can be stored on a small amount of memory. Thus, we focus on the asymptotic regime, where the number of samples $N$ is large. We characterize the optimal performance in the asymptotic regime, which is associated with the maximum likelihood estimator (MLE) for the deterministic setup and the joint maximum \textit{a-posteriori} probability-maximum likelihood estimator (JMAP-MLE) for the hybrid setup and establish the $1$-bit loss with an unknown quantization level.
\section{System Model}
\subsection{Ideal Receiver ($\infty$-bit resolution)}
For the analysis we consider the problem of pilot-based channel estimation. Therefore, the digital signal model of the ideal receive system with infinite resolution is given by
\begin{align}
\ve{y} 
&= \zeta\ve{x}+\ve{\eta},
\label{def:signal:model}
\end{align}
where $\ve{x}\in\fieldR^N$ is a pilot signal of known structure, $\zeta\in\fieldR$ characterizes the channel attenuation, and $\ve{\eta}\in\fieldR^N$ is unit-variance white Gaussian noise. Therefore, given the parameter $\zeta$, the received signal $\ve{y}\in\fieldR^N$ follows the conditional probability law $\ve{y}|\zeta\sim\mathcal{N}( \zeta\ve{x},\ve{I})$, where $\ve{0}\in\fieldR^N$ denotes the all-zeros vector and $\ve{I}\in\fieldR^{N\times N}$ is the identity matrix. We assume a quasi-ergodic phase-shift keying (BPSK) transmitter and a synchronized receiver such that $\ve{x}\in\{-1,1\}^N$ where
\begin{align}
\label{eq:restrictions}
\frac{1}{N}\sum_{n=1}^{N}x_n&=0.
\end{align} 
\subsection{Low-Complexity Receiver ($1$-bit resolution)}
The low-complexity receiver under investigation is equipped with a $1$-bit ADC which provides the digital signal
\begin{align}\label{system:model:sign}
\ve{r}&=\sign{\ve{y}-\alpha\ve{1}}=\sign{ \zeta\ve{x}+\ve{\eta}-\alpha\ve{1}},
\end{align}
where $\sign{\ve{x}}$ is the element-wise signum function
\begin{align}
\left[\sign{\ve{x}}\right]_{n}=
\begin{cases}
+1& \text{if } x_n \geq 0\\
-1& \text{if } x_n < 0
\end{cases},\ n=1,\ldots,N,
\end{align}
$\ve{1}\in\fieldR^N$ denotes the all-ones vector and $\alpha\in\fieldR$ forms the unknown threshold level. The conditional probability density function (PDF) of the resulting binary received signal is
\begin{align}
\label{eq:condpdf}
p_{\ve{r}}(\ve{r};\zeta,\alpha) &= \prod_{n=1}^{N} \qfunc{ r_n(\alpha- \zeta x_n)},
\end{align}
with $\qfunc{x}$ being the Q-function
\begin{align}
\qfunc{x}=\frac{1}{\sqrt{2\pi}} \int_{x}^{\infty} \exp{\Big(-\frac{z^2}{2}\Big)} {\rm d} z.
\end{align}
The task of the receivers is to calculate the estimates $\hat{\zeta}_{\ve{y}}(\ve{y})$ and $\hat{\zeta}_{\ve{r}}(\ve{r})$ by using the received signal $\ve{y}$ or $\ve{r}$, respectively.
\subsection{Deterministic Approach}
Under this approach the ideal receiver treats $\zeta$ as deterministic unknown, such that the asymptotically optimum unbiased estimator is the maximum likelihood estimator (MLE), given by
\begin{align}
\hat{\zeta}_{\ve{y}}(\ve{y})&\triangleq\arg \max_{\zeta\in \fieldR} p_{\ve{y}}(\ve{y};\zeta)\notag\\
&=\arg \max_{\zeta\in \fieldR} \sum_{n=1}^{N} \ln p_{y_n}(y_n;\zeta),
\end{align}
with the corresponding error
\begin{align}\label{eq:FMSEy}
\text{MSE}_{\ve{y}}(\zeta)&=\exdi{\ve{y};\zeta}{\big(\hat{\zeta}_{\ve{y}}(\ve{y})-\zeta\big)^2}.
\end{align}
The $1$-bit receiver considers both $\zeta$ and the threshold $\alpha$ as deterministic unknown. The MLE is based on joint estimation of both the parameter $\zeta$ and the threshold $\alpha$, such that
\begin{align}
\begin{bmatrix}\hat{\zeta}_{\ve{r}}(\ve{r}) &\hat{\alpha}_{\ve{r}}(\ve{r})\end{bmatrix}^{\T}&\triangleq\arg \max_{\zeta, \alpha \in \fieldR } p_{\ve{r}}(\ve{r};\zeta,\alpha)\notag\\
&=\arg \max_{\zeta, \alpha \in \fieldR } \sum_{n=1}^{N} \ln p_{r_n}(r_n;\zeta,\alpha),
\end{align}
with the corresponding error
\begin{align}\label{eq:FMSEr}
\text{MSE}_{\ve{r}}(\zeta,\alpha)&\triangleq\exdi{\ve{r};\zeta,\alpha}{\big(\hat{\zeta}_{\ve{r}}(\ve{r})-\zeta\big)^2}.
\end{align}
\subsection{Hybrid Approach}
The second approach considers the parameter $\zeta$ to be modeled as a random variable with a prior PDF $p_\zeta(\zeta)$. The received signal $\ve{y}$ and the parameter of interest $\zeta$ follow the joint PDF $p_{\ve{y},\zeta}(\ve{y},\zeta)$. The asymptotically optimum estimator is the maximum \textit{a-posteriori} probability (MAP) estimator
\begin{align}\label{eq:MAPy}
\hat{\zeta}_{\ve{y}}(\ve{y})&\triangleq\arg \max_{\zeta\in \fieldR} p_{\ve{y},\zeta}(\ve{y},\zeta)\notag\\
&=\arg \max_{\zeta\in \fieldR} \left(\ln p_{\ve{y}|\zeta}(\ve{y}|\zeta)+\ln p_{\zeta}(\zeta)\right),
\end{align}
where the last equality stems from Bayes law. The corresponding error is defined as
\begin{align}
\label{eq:HMSEy}
\text{MSE}_{\ve{y}}&\triangleq\exdi{\ve{y},\zeta}{\big(\hat{\zeta}_{\ve{y}}(\ve{y})-\zeta\big)^2}.
\end{align}
The $1$-bit receiver treats $\zeta$ as random while the threshold $\alpha$ remains deterministic unknown. The received signal $\ve{r}$ and the parameter $\zeta$ follow the joint PDF $p_{\ve{r},\zeta}(\ve{r},\zeta;\alpha)$. For the $1$-bit receiver, the asymptotically optimum estimator \cite{Bar15_2} in the MSE sense is the JMAP-MLE \cite{Yeredor00}, given by  
\begin{align}\label{eq:JMAPML}
\begin{bmatrix}\hat{\zeta}_{\ve{r}}(\ve{r}) &\hat{\alpha}_{\ve{r}}(\ve{r})\end{bmatrix}^{\T}&\triangleq\arg \max_{\zeta,\alpha\in \fieldR} p_{\ve{r},\zeta}(\ve{r},\zeta;\alpha)\notag\\
&=\arg \max_{\zeta, \alpha \in \fieldR } \left(\ln p_{\ve{r}|\zeta}(\ve{r}|\zeta;\alpha)+ \ln p_{\zeta}(\zeta)\right).
\end{align}
The corresponding error is defined as
\begin{align}
\label{eq:HMSEr}
\text{MSE}_{\ve{r}}(\alpha)&\triangleq\exdi{\ve{r},\zeta;\alpha}{\big(\hat{\zeta}_{\ve{r}}(\ve{r})-\zeta\big)^2}.
\end{align}
\section{Performance Analysis}
In this section, the expressions for the MSEs in \eqref{eq:FMSEy}, \eqref{eq:FMSEr}, \eqref{eq:HMSEy}, and \eqref{eq:HMSEr} are evaluated. Note that due to the possibility of high sampling rates with $1$-bit ADC, the focus is on the asymptotic regime, where the number of samples $N$ is large. Thus, the CRLB and its expected version utilized in the sequel are used as valid approximations to the subjected MSEs.

\subsection{Deterministic Approach - Hard-limiting Loss}
With the ideal receiver and estimation by the MLE, the MSE can be approximated asymptotically by the CRLB \cite{Rao45,Cram46}
\begin{align}
\label{eq:CRLB}
\text{MSE}_{\ve{y}}(\zeta)\asympequal F_{y}^{-1}(\zeta),
\end{align}
where with \eqref{def:signal:model} the Fisher information (FI) \cite{Kay93} is
\begin{align}
F_{y}(\zeta)&=\exdi{\ve{y};\zeta}{ \bigg(\frac{\partial \ln p_{\ve{y}}(\ve{y};\zeta) }{\partial \zeta} \bigg)^2}=
\sum_{n=1}^{N} x_n^2=N.
\end{align}
For the $1$-bit receiver, the estimation of the threshold $\hat{\alpha}_{\ve{r}}(\ve{r})$ has an effect onto the inference of the attenuation parameter $\hat{\zeta}_{\ve{r}}(\ve{r})$. The corresponding CRLB for the estimator $\hat{\zeta}_{\ve{r}}(\ve{r})$ is
\begin{align}
\label{eq:CRLB:1bit}
\text{MSE}_{\ve{r}}(\zeta,\alpha)\asympequal \frac{F_{r,\alpha \alpha}(\zeta,\alpha)}{F_{r,\zeta \zeta}(\zeta,\alpha)F_{r,\alpha \alpha}(\zeta,\alpha)-F_{r,\zeta \alpha}^2(\zeta,\alpha)}.
\end{align}
The required FIs are given by
\begin{align}\label{info:fisher:zeta:zeta}
F_{r,\zeta \zeta}(\zeta,\alpha) &=\exdi{\ve{r};\zeta,\alpha}{ \bigg(\frac{\partial \ln p_{\ve{r}}(\ve{r};\zeta,\alpha) }{\partial \zeta} \bigg)^2} \notag\\
&=\sum_{n=1}^{N}\exdi{r_n;\zeta,\alpha}{ \frac{ x_n^2\expb{ -(\alpha-\zeta x_n)^2 } }{ {2\pi}\qfuncptwo{r_n(\alpha- \zeta x_n)} }   }\notag\\
&=\sum_{n=1}^{N} \frac{ x_n^2\expb{ -(\alpha-\zeta x_n)^2 }  }{ {2\pi} \big(\qfunc{\alpha- \zeta x_n}-\qfuncptwo{\alpha- \zeta x_n}\big)}\notag\\
&= \frac{N}{2}\big(\phi_{+}(\zeta, \alpha)+\phi_{-}(\zeta, \alpha)\big),
\end{align}
where the third equality stems from \eqref{eq:condpdf}, such that 
\begin{align}
&\exdi{r_n;\zeta,\alpha}{ \frac{1}{ \qfuncptwo{r_n(\alpha- \zeta x_n)} } }
=\sum_{r_n=\pm 1}{ \frac{\qfunc{ r_n(\alpha- \zeta x_n)}}{ \qfuncptwo{r_n(\alpha- \zeta x_n)} } }\nonumber\\
&=\frac{1}{ \qfunc{\alpha- \zeta x_n}-\qfuncptwo{\alpha- \zeta x_n}}
\end{align}
and the last equality stems from the BPSK modulation of $\{x_n\}_{n=1}^N$ and the definition
\begin{align}
\phi_{\pm}(\zeta, \alpha) &= \frac{ \expb{ -(\alpha \pm \zeta)^2 } }{ {2\pi} \big(\qfunc{\alpha \pm \zeta}-\qfuncptwo{\alpha \pm \zeta}\big)}.
\end{align} 
In the same manner,
\begin{align}
F_{r,\alpha \alpha}(\zeta,\alpha) &=\exdi{\ve{r};\zeta,\alpha}{ \bigg(\frac{\partial \ln p_{\ve{r}}(\ve{r};\zeta,\alpha) }{\partial \alpha} \bigg)^2} \notag\\
&=\sum_{n=1}^{N}\exdi{r_n;\zeta,\alpha}{ \frac{ \expb{ -(\alpha-\zeta x_n)^2 } }{ {2\pi}\qfuncptwo{r_n(\alpha- \zeta x_n)} }  }\notag\\
&= \frac{N}{2}\big(\phi_{+}(\zeta, \alpha)+\phi_{-}(\zeta, \alpha)\big),\label{info:fisher:alpha:alpha}\\
F_{r,\zeta \alpha}(\zeta,\alpha)&=\exdi{\ve{r};\zeta,\alpha}{ \frac{\partial \ln p_{\ve{r}}(\ve{r};\zeta,\alpha) }{\partial \zeta} \frac{\partial \ln p_{\ve{r}}(\ve{r};\zeta,\alpha) }{\partial \alpha} } \notag\\
&=-\sum_{n=1}^{N}\exdi{r_n;\zeta,\alpha}{ \frac{ x_n\expb{ -(\alpha-\zeta x_n)^2 } }{ {2\pi}\qfuncptwo{r_n(\alpha- \zeta x_n)} } }\notag\\
&=\frac{N}{2}\big(\phi_{+}(\zeta, \alpha)-\phi_{-}(\zeta, \alpha)\big).\label{info:fisher:zeta:alpha}
\end{align}
Note that if the quantization level is known to the $1$-bit receiver, the performance is limited to
\begin{align}\label{eq:CRLB:1bit:star}
\text{MSE}^{\star}_{r}(\zeta,\alpha)\asympequal F^{-1}_{r,\zeta \zeta}(\zeta,\alpha).
\end{align}
To characterize the information loss introduced by the hard-limiter, we define the quantization loss via two MSE ratios
\begin{align}
\chi(\zeta,\alpha)&\triangleq\frac{\text{MSE}_{\ve{y}}(\zeta)}{\text{MSE}_{\ve{r}}(\zeta,\alpha)}\notag\\
&\asympequal\frac{F_{r,\zeta \zeta}(\zeta,\alpha)F_{r,\alpha \alpha}(\zeta,\alpha)-F_{r,\zeta \alpha}^2(\zeta,\alpha)}{F_{r,\alpha \alpha}(\zeta,\alpha)F_{y,\zeta \zeta}(\zeta)}\notag\\
&=2 \frac{ \phi_{+}(\zeta, \alpha)\phi_{-}(\zeta, \alpha)\big)}{ \phi_{+}(\zeta, \alpha)+\phi_{-}(\zeta, \alpha)}\label{loss:det:det},\\
\chi^{\star}(\zeta,\alpha)&\triangleq\frac{\text{MSE}_{\ve{y}}(\zeta)}{\text{MSE}^{\star}_{r}(\zeta,\alpha)}\asympequal\frac{F_{r,\zeta \zeta}(\zeta,\alpha)}{F_{y,\zeta \zeta}(\zeta)}\notag\\
&=\frac{1}{2}\big(\phi_{+}(\zeta, \alpha)+\phi_{-}(\zeta, \alpha)\big)\label{loss:det:det:knownt},
\end{align}
which are associated with the ideal receiver and with a known quantization level $\alpha$ at the $1$-bit receiver, respectively. Fig. \ref{HardLimLoss_DD} shows the performance loss \eqref{loss:det:det} for different SNR levels in solid lines, where we use the convention $\text{SNR}=\zeta^2$.
\pgfplotsset{legend style={rounded corners=4pt,nodes=right}}
\begin{figure}[!htbp]
\begin{tikzpicture}[scale=1.0]

  	\begin{axis}[ylabel=$\chi(\zeta\text{,}\alpha)\text{ in dB}$,
  			xlabel=$\alpha$,
			grid,
			ymin=-5,
			ymax=-1.5,
			xmin=0,
			xmax=1,
			legend pos= south west]
			
    			\addplot[black, style=solid, line width=0.75pt,smooth, every mark/.append style={solid, fill=black}, mark=diamond*, mark repeat=2] table[x index=0, y index=1]{HarLimLoss_DD.txt};
			\addlegendentry{$\text{SNR}=-25\text{ dB}$};
			
			\addplot[green, style=solid, line width=0.75pt,smooth, every mark/.append style={solid}, mark=otimes*, mark repeat=2] table[x index=0, y index=2]{HarLimLoss_DD.txt};
			\addlegendentry{$\text{SNR}=-10\text{ dB}$};
			
			\addplot[blue, style=solid, line width=0.75pt,smooth, every mark/.append style={solid}, mark=square*, mark repeat=2] table[x index=0, y index=3]{HarLimLoss_DD.txt};
			\addlegendentry{$\text{SNR}=-5.0\text{ dB}$};
			
			\addplot[red, style=solid, line width=0.75pt,smooth, every mark/.append style={solid}, mark=triangle*, mark repeat=2] table[x index=0, y index=4]{HarLimLoss_DD.txt};
			\addlegendentry{$\text{SNR}=-2.5\text{ dB}$};
			
			\addplot[black, style=dashed, line width=0.75pt,smooth, every mark/.append style={solid, fill=black}, mark=diamond*, mark repeat=2] table[x index=0, y index=1]{HarLimLoss_DDK.txt};
			\addplot[green, style=dashed, line width=0.75pt,smooth, every mark/.append style={solid}, mark=otimes*, mark repeat=2] table[x index=0, y index=2]{HarLimLoss_DDK.txt};
			\addplot[blue, style=dashed, line width=0.75pt,smooth, every mark/.append style={solid}, mark=square*, mark repeat=2] table[x index=0, y index=3]{HarLimLoss_DDK.txt};
			\addplot[red, style=dashed, line width=0.75pt,smooth, every mark/.append style={solid}, mark=triangle*, mark repeat=2] table[x index=0, y index=4]{HarLimLoss_DDK.txt};

	\end{axis}

\end{tikzpicture}
\caption{Frequentist - $\chi$ (solid) and $\chi^{\star}$ (dashed)}
\label{HardLimLoss_DD}
\end{figure}
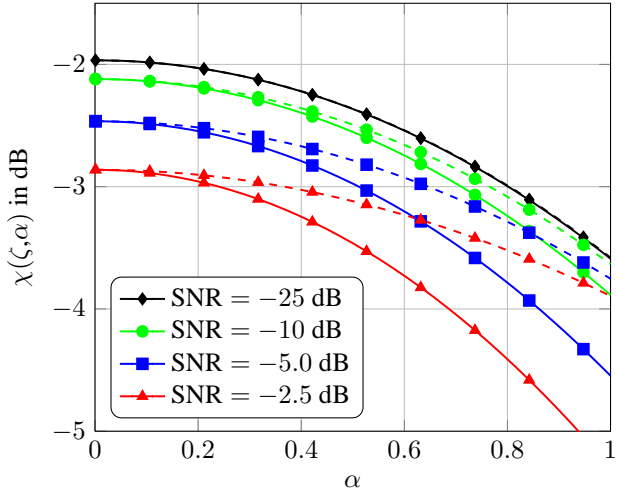
Note that the loss is symmetric for negative $\alpha$ or $\zeta$. The results show that for the considered application a quantization level $\alpha$ close to zero is in general  preferable and that the performance gap increases with the SNR. Additionally, with dashed lines, Fig. \ref{HardLimLoss_DD} shows the alternative loss \eqref{loss:det:det:knownt}. While in the low SNR regime the estimation of $\alpha$ has no effect onto the estimation of $\zeta$, the situation changes within the medium SNR regime. Here the fact that the threshold is unknown can have a significant effect, in particular when $\alpha$ is not close to zero. 
\subsection{Hybrid Approach - Hard-limiting Loss}
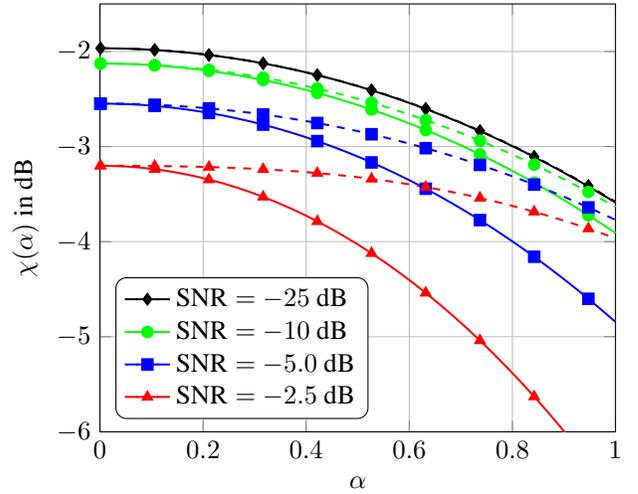
\begin{figure}[!htbp]
\begin{tikzpicture}[scale=1.0]

  	\begin{axis}[ylabel=$\chi(\alpha)\text{ in dB}$,
  			xlabel=$\alpha$,
			grid,
			ymin=-6,
			ymax=-1.5,
			xmin=0,
			xmax=1,
			legend pos= south west]
			
    			\addplot[black, style=solid, line width=0.75pt,smooth, every mark/.append style={solid, fill=black}, mark=diamond*, mark repeat=2] table[x index=0, y index=1]{HarLimLoss_RD_E.txt};
			\addlegendentry{$\text{SNR}=-25\text{ dB}$};
			
			\addplot[green, style=solid, line width=0.75pt,smooth, every mark/.append style={solid}, mark=otimes*, mark repeat=2] table[x index=0, y index=2]{HarLimLoss_RD_E.txt};
			\addlegendentry{$\text{SNR}=-10\text{ dB}$};
			
			\addplot[blue, style=solid, line width=0.75pt,smooth, every mark/.append style={solid}, mark=square*, mark repeat=2] table[x index=0, y index=3]{HarLimLoss_RD_E.txt};
			\addlegendentry{$\text{SNR}=-5.0\text{ dB}$};
			
			\addplot[red, style=solid, line width=0.75pt,smooth, every mark/.append style={solid}, mark=triangle*, mark repeat=2] table[x index=0, y index=4]{HarLimLoss_RD_E.txt};
			\addlegendentry{$\text{SNR}=-2.5\text{ dB}$};

    			\addplot[black, style=dashed, line width=0.75pt,smooth, every mark/.append style={solid}, mark=diamond*, mark repeat=2] table[x index=0, y index=1]{HarLimLoss_RD_EK.txt};
			\addplot[green, style=dashed, line width=0.75pt,smooth, every mark/.append style={solid}, mark=otimes*, mark repeat=2] table[x index=0, y index=2]{HarLimLoss_RD_EK.txt};
			\addplot[blue, style=dashed, line width=0.75pt,smooth, every mark/.append style={solid}, mark=square*, mark repeat=2] table[x index=0, y index=3]{HarLimLoss_RD_EK.txt};
			\addplot[red, style=dashed, line width=0.75pt,smooth, every mark/.append style={solid}, mark=triangle*, mark repeat=2] table[x index=0, y index=4]{HarLimLoss_RD_EK.txt};

	\end{axis}

\end{tikzpicture}
\caption{Hybrid - $\chi$ (solid) and $\chi^{\star}$ (dashed)}
\label{HardLimLoss_RD}
\end{figure}

In the case of a random channel parameter, with the ideal receiver, the asymptotic performance of the MAP estimator can be characterized using the expected CRLB \cite[p. 6]{Trees07}
\begin{align}
\label{eq:ECRLB}
\text{MSE}_{\ve{y}} \asympequal \exdi{\zeta}{F_{y}^{-1}(\zeta)}=\frac{1}{N},
\end{align}
where the last equality stems from taking the expectation of \eqref{eq:CRLB} with respect to $\zeta$. For the $1$-bit receiver, according to \cite{Ibragimov81}, as the number of measurements increases, the JMAP-MLE in \eqref{eq:JMAPML}, coincides (in the sense of convergence in probability) with the MLE, given by
\begin{equation}
\begin{bmatrix}\hat{\zeta}^{(\text{ML})}_{\ve{r}}(\ve{r}) & \hat{\alpha}^{(\text{ML})}_{\ve{r}}(\ve{r})\end{bmatrix}^{\T}=\arg \max_{\zeta, \alpha \in \fieldR } \ln p_{\ve{r}|\zeta}(\ve{r}|\zeta;\alpha).
\end{equation}
The sequence of MLEs as a function of the number of measurements is asymptotically uniformly integrable \cite{Jeganathan82}. Thus,
\begin{align}
\label{eq:EHCRLBPromo}
&\lim\limits_{N\to\infty}\text{MSE}_{\ve{r}}(\alpha)=\lim\limits_{N\to\infty}\exdi{\ve{r},\zeta;\alpha}{\big(\hat{\zeta}^{(\text{ML})}_{\ve{r}}(\ve{r})-\zeta\big)^2}\nonumber\\
&=\lim\limits_{N\to\infty}\E_{\zeta} \Big\{{\exdi{\ve{r}|\zeta;\alpha}{\big(\hat{\zeta}^{(\text{ML})}_{\ve{r}}(\ve{r})-\zeta\big)^2}}\Big\}\nonumber
\\
&=\lim\limits_{N\to\infty}\exdi{\zeta}{\frac{F_{r,\alpha \alpha}(\zeta,\alpha)}{F_{r,\zeta \zeta}(\zeta,\alpha)F_{r,\alpha \alpha}(\zeta,\alpha)-F_{r,\zeta \alpha}^2(\zeta,\alpha)}},
\end{align}
where the last equality stems from taking the expectation of \eqref{eq:CRLB:1bit}. Hence, by using \eqref{info:fisher:zeta:zeta}, \eqref{info:fisher:alpha:alpha}, and \eqref{info:fisher:zeta:alpha}, one obtains
\begin{align}
\label{eq:EHCRLB}
&\text{MSE}_{\ve{r}}(\alpha)\asympequal\exdi{\zeta}{\frac{F_{r,\alpha \alpha}(\zeta,\alpha)}{F_{r,\zeta \zeta}(\zeta,\alpha)F_{r,\alpha \alpha}(\zeta,\alpha)-F_{r,\zeta \alpha}^2(\zeta,\alpha)}}\notag\\
&=\exdi{\zeta}{ \frac{\frac{2}{N}\big(\phi_{+}(\zeta, \alpha)+\phi_{-}(\zeta, \alpha)\big)}{\big(\phi_{+}(\zeta, \alpha)+\phi_{-}(\zeta, \alpha)\big)^2-\big(\phi_{+}(\zeta, \alpha)-\phi_{-}(\zeta, \alpha)\big)^2} }\notag\\
&= \frac{1}{2N}\Bigg(\exdi{\zeta}{ \frac{1}{ \phi_{-}(\zeta, \alpha) }}+\exdi{\zeta}{ \frac{1}{ \phi_{+}(\zeta, \alpha) }}\Bigg)= \frac{1}{N}\Psi_H,
\end{align}
where due to symmetry considerations, we define
\begin{align}
\Psi_H\triangleq\exdi{\zeta}{ \frac{1}{ \phi_{-}(\zeta, \alpha) }}=\exdi{\zeta}{ \frac{1}{ \phi_{+}(\zeta, \alpha) }}.
\end{align}
The quantization losses are given by
\begin{align}
\chi(\alpha)&=\frac{\text{MSE}_{\ve{y}}}{\text{MSE}_{\ve{r}}(\alpha)}\asympequal \Psi^{-1}_H\label{loss:rand:det:exp},\\
\chi^{\star}(\alpha)&\triangleq\frac{\text{MSE}_{\ve{y}}}{\text{MSE}^{\star}_{\ve{r}}(\alpha)}\asympequal\frac{\exdi{\zeta}{F_{y}^{-1}(\zeta)}}{\exdi{\zeta}{F^{-1}_{r,\zeta \zeta}(\zeta,\alpha)}}\notag\\
&=\frac{1}{2\exdi{\zeta}{ \frac{1}{\phi_{+}(\zeta, \alpha)+\phi_{-}(\zeta, \alpha)} }}\label{loss:rand:det:knownt}.
\end{align}
Fig. \ref{HardLimLoss_RD} shows the performance loss \eqref{loss:rand:det:exp} with solid lines. In this scenario $\text{SNR}=\sigma_{\zeta}^2$ as the noise $\eta_n\sim\mathcal{N}(0,1)$.
It can be observed that the quantization loss increases sharply when the hard-limiting level $\alpha$ deviates from zero. Additionally, with dashed lines, Fig. \ref{HardLimLoss_RD} shows the alternative performance loss \eqref{loss:rand:det:knownt}. Like in the deterministic setup, the unknown quantization introduces an additional loss which becomes small for low SNR and when the unknown threshold is close to zero.
\section{Conclusion}
We have analyzed the performance of a $1$-bit receiver with respect to the task of channel parameter estimation when the quantization offset is unknown. In this situation the receiver has to estimate the quantization level which in general has an influence on the quality of the estimation of the channel parameter. In the low SNR regime and when the quantization offset is close to zero this effect vanishes. This confirms that $1$-bit ADCs are an interesting option for low SNR applications while for the medium SNR regime signal processing with $1$-bit ADCs requires careful hardware design.

\end{document}